\documentclass{pasj00}
\usepackage{rotating}

\begin{document}
\SetRunningHead{Iono et al.}{Dense Gas in VV114}

\title{Active Galactic Nucleus and Extended Starbursts in a Mid-stage Merger VV114}

\author{Daisuke \textsc{Iono}\altaffilmark{1,2},
Toshiki \textsc{Saito}\altaffilmark{1,3},
Min S. \textsc{Yun}\altaffilmark{4},
Ryohei \textsc{Kawabe}\altaffilmark{1,5},
Daniel \textsc{Espada}\altaffilmark{1,5},
Yoshiaki \textsc{Hagiwara}\altaffilmark{1,2},
Masatoshi \textsc{Imanishi}\altaffilmark{1,2},
Takuma \textsc{Izumi}\altaffilmark{6},
Kotaro \textsc{Kohno}\altaffilmark{6,7},
Kentaro \textsc{Motohara}\altaffilmark{6},
Koichiro \textsc{Nakanishi}\altaffilmark{2,5},
Hajime \textsc{Sugai}\altaffilmark{8},
Ken \textsc{Tateuchi}\altaffilmark{6},
Yoichi \textsc{Tamura}\altaffilmark{6},
Junko \textsc{Ueda}\altaffilmark{1,3,9},
Yuzuru \textsc{Yoshii}\altaffilmark{6}
} %
\altaffiltext{1}{National Astronomical Observatory of Japan, 
2-21-1 Osawa, Mitaka, Tokyo 181-8588}
\altaffiltext{2}{The Graduate University for Advanced Studies 
(SOKENDAI), 2-21-1 Osawa, Mitaka, Tokyo 181-0015}
\altaffiltext{3}{Department of Astronomy, School of Science, The University of Tokyo, 7-3-1 Hongo, Bunkyo-ku, Tokyo 133-0033}
\altaffiltext{4}{Department of Astronomy, University of Massachusetts, Amherst, MA 01003}
\altaffiltext{5}{Joint ALMA Observatory, Alonso de Cordova 3107, Vitacura, Santiago 763-0355, Chile}
\altaffiltext{6}{Institute of Astronomy, University of Tokyo, 2-21-1 Osawa, Mitaka, Tokyo 181-0015}
\altaffiltext{7}{Research Center for the Early Universe (WPI), University of Tokyo, 7-3-1 Hongo, Bunkyo, Tokyo 113-0033, Japan}
\altaffiltext{8}{Kavli Institute for the Physics and Mathematics of the Universe, The Univ. of Tokyo}
\altaffiltext{9}{Harvard-Smithsonian Center for Astrophysics, 60 Garden Street, Cambridge, MA 02138}

%

\KeyWords{telescopes --- galaxies:evolution --- galaxies:starburst  --- galaxies:interactions} 

\maketitle

\begin{abstract}
High resolution ($\sim 0 \farcs4$) 
Atacama Large Millimeter/submillimeter Array (ALMA) Cycle 0 observations of 
HCO$^+$(4--3) and HCN~(4--3) toward a mid-stage infrared bright merger 
VV~114 have revealed compact nuclear ($< 200$~pc) 
and extended ($\sim 3 - 4$~kpc) dense gas distribution across 
the eastern part of the galaxy pair.  
We find significant  
enhancement of HCN~(4--3) emission in an unresolved compact and 
broad (290~km~s$^{-1}$) component
found in the eastern nucleus of VV114, 
and we suggest dense gas associated with 
the surrounding material around an Active Galactic Nucleus (AGN), with a 
mass upper limit of $\lesssim 4 \times 10^8$~M$_{\odot}$.  
The extended dense gas is distributed along 
a filamentary structure with resolved dense gas concentrations 
($\sim 230$~pc; $\sim10^6$~M$_{\odot}$) separated 
by a mean projected distance of $\sim 600$~pc, 
many of which are generally consistent
with the location of star formation traced in Pa$\alpha$ emission.  
Radiative transfer calculations suggest moderately dense
($n_{\rm H_2} = 10^5$ -- $10^6$~cm$^{-3}$) gas 
averaged over the entire emission region.  
These new ALMA observations demonstrate 
the strength of the dense gas tracers in identifying both the AGN 
and star formation activity in a galaxy merger, 
even in the most dust enshrouded environments  
in the local universe.
\end{abstract}

\section{Introduction}

Cosmological simulations have clearly established that 
galaxy collisions and mergers play major roles in the 
formation and evolution of galaxies by triggering a rapid 
mass build-up (e.g., \cite{cole00}).  
High-resolution major merger simulations have shown that 
the star formation physics is more 
dominated by mass fragmentation and turbulent motion across the 
merging disks, forming massive clumps of dense gas clouds 
($M_{\rm gas} = 10^{6-8}$~M$_\odot$) and triggering star 
formation across the galaxy disks \citep{teyssier10}, or in 
a dense filamentary structure along the merging interface 
\citep{saitoh09}. 
In some cases, radial streaming can efficiently feed the gas 
to the central black hole, possibly triggering an 'AGN phase' 
during the course of the galaxy merger evolution 
(e.g. \cite{hopkins06}).




An important observational test is to map the dense gas tracers 
in merging U/LIRGs (Ultra Luminous Infrared Galaxies) since 
they show high degree of starburst activity, some of which 
harbor AGNs in their centers \citep{sanders96}.  
The HCN~(4--3) and HCO$^+$(4--3) emission, whose critical 
densities are $n_{\rm crit} \sim 2 \times 10^7$~cm$^{-3}$ and  
$n_{\rm crit} \sim 4 \times 10^6$~cm$^{-3}$ \citep{meijerink07} 
respectively, are both 
reliable dense gas tracers and now readily  
accessible at sub arcsecond angular resolution
with the advent of the Atacama Large Millimeter/Submillimeter Array (ALMA).
Here we present ALMA Cycle 0 HCN~(4--3) and HCO$^+$(4--3) observations of an 
IR-bright galaxy VV114 with the primary goal to 
study the distribution of dense gas during the critical stage when 
the two gas rich galaxies collide and merge.


VV114 is a gas-rich ($M_{\rm H_2} = 5.1 \times 10^{10}~\rm M_{\odot}$; 
\cite{yun94}, \cite{iono04} and \cite{wilson08})
interacting pair with high-infrared luminosity 
($L_{IR} = 4.0 \times 10^{11}~\rm L_{\odot}$) located at D = 86~Mpc
($1\arcsec = 0.4$~kpc). 
It consists of two optical galaxies (VV114E and VV114W) with a projected
separation of 6~kpc (Figure~1). 
Evidence for wide-spread star formation activity 
and shocks across the entire system is found in the UV, optical, and mid-IR 
(\cite{alonsoherrero02}, 
\cite{goldader02},
\cite{rich11}). 
A dust-obscured AGN in VV114E is also suggested (but not conclusive)
from NIR \citep{lefloch02, imanishi07} 
and X-ray observations \citep{grimes06}, 
suggesting that both starburst and AGN activities might have been 
triggered by the ongoing merger.

\section{ALMA Observations}

\begin{figure*}
  \begin{center}
    \FigureFile(160mm,40mm){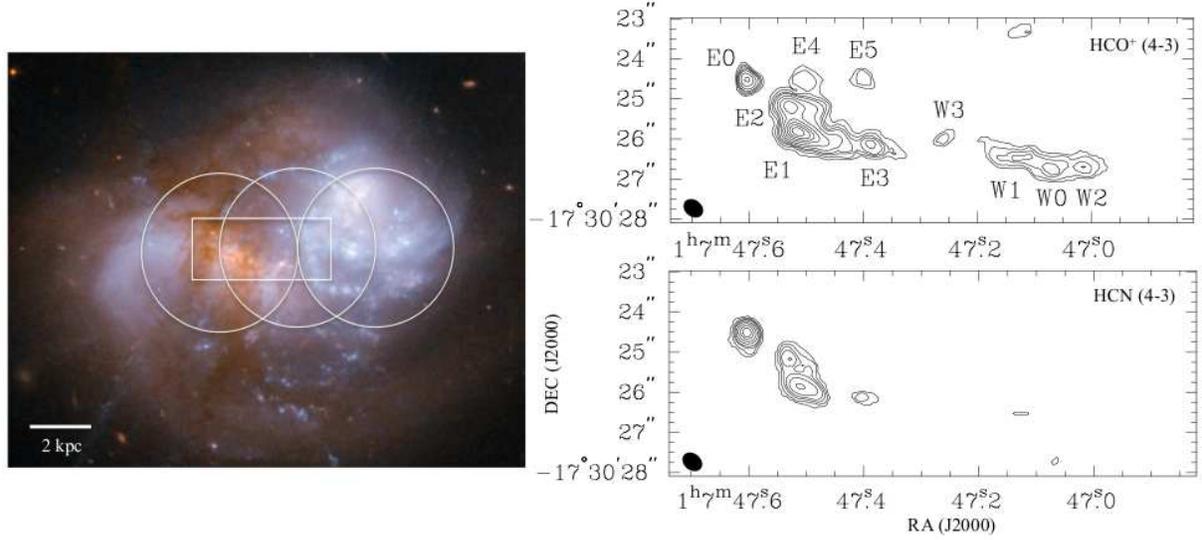}
  \end{center}
  \caption{ (left) HST ACS image of VV114 overlaid with the approximate regions of the panels shown on the right, and the approximate field of views of the ALMA 3-point mosaic. (Credit: NASA, ESA, the Hubble Heritage (STScI/AURA)-ESA/Hubble Collaboration, and A. Evans (University of Virginia, Charlottesville/NRAO/Stony Brook University). (right) The distribution of HCO$^+$(4--3) (top) and HCN~(4--3) (bottom) in VV~114.  The contours are; 0.05, 0.15, 0.25, 0.35, 0.55, 0.85, 1.25, 1.65, 2.05, 2.45 Jy~km~s$^{-1}$.}
\end{figure*}

The HCN~(4--3) ($\nu_{rest} = 354.505$~GHz) 
and HCO$^+$(4--3) ($\nu_{rest} = 356.734$~GHz) observations toward 
VV114 were obtained on June 1 -- 3, 2012 
during the Cycle 0 program of ALMA using
the extended configuration.
The digital correlator was configured with 0.488~MHz resolution for the 
spectral window that contains the emission lines.
Absolute flux calibration was performed using Uranus, 
J1924-292 was used for bandpass calibration,  
and the time dependent gain calibration was performed using 
J0132-169 (6 degrees away from VV~114).
The total on source time was 86 minutes. 

We used the delivered calibrated  
data product and CLEANed the image down to 1.5 sigma using the ALMA 
data reduction package CASA.
Channel maps with 30 km~s$^{-1}$ velocity resolution were made, 
with a synthesized beam size of $0\farcs5 \times 0\farcs4$ 
(PA = 52 degrees)(equivalent to $200 \times 160$~pc).  
The rms noise level was 0.9 mJy~beam$^{-1}$ for the robust=0.5 maps. 
The continuum was subtracted using all of the 
line-free channels in the bandpass.  

\begin{figure*}
  \begin{center}
    \FigureFile(160mm,20mm){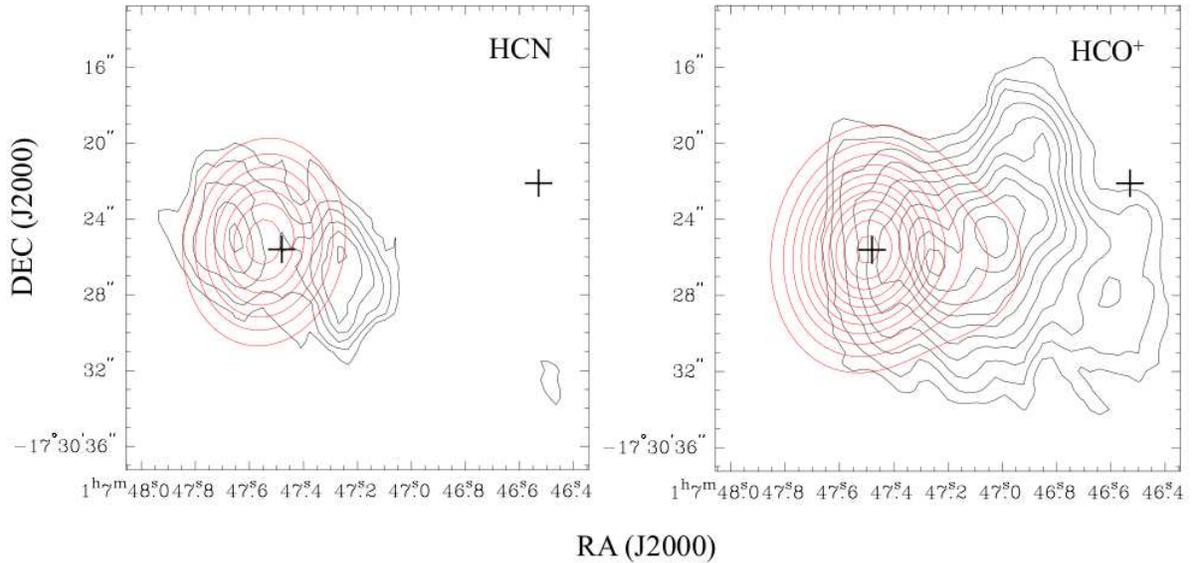}

  \end{center}
  \caption{(left) The HCN~(1--0) emission (\cite{imanishi07}; in dark contours) is compared with
the HCN~(4--3) emission convolved to the NMA resolution (in red contours).  
The contour levels for HCN~(1--0) 
are the same as \citet{imanishi07} and the HCN~(4--3) contours are 
$(2.8 - 4.8) \times 10^{-2}$ (in steps of $0.2 \times 10^{-2}$)
Jy/beam~km~s$^{-1}$.  The crosses indicate the locations
of the near-infrared peaks shown in Figure~2 of \citet{imanishi07}. 
(right) Similar to $left$ but for the HCO$^+$ emission.
The HCN~(4--3) contours are 
$(1.0 - 1.7) \times 10^{-2}$ (in steps of $0.1 \times 10^{-2}$)
Jy/beam~km~s$^{-1}$.  
}
\end{figure*}

\section{Distribution of  HCN~(4--3)  and HCO$^+$(4--3)}

The HCN~(4--3) and HCO$^+$(4--3) integrated intensity maps 
are presented in Figure~1.  
While the HCN~(4--3) emission is only seen near the eastern nucleus of 
VV114 and resolved into four peaks, the 
HCO$^+$(4--3) emission is more extended and 
has at least 10 peaks in the integrated intensity map.  
The total integrated intensity of HCO$^+$(4--3) and HCN~(4--3) are 
$15.3 \pm 0.4$ Jy~km~s$^{-1}$ and $4.4 \pm 0.2$ Jy~km~s$^{-1}$, respectively.
The higher HCO$^+$(4--3) flux observed by the SMA ($\geq 17 \pm 2$~mJy; \cite{wilson08})
using a $2\farcs8 \times 2\farcs0$ beam is likely attributed to 
missing flux by the ALMA observation.
We show a direct comparison between the J=4--3 (this work) 
and J=1--0 (taken at the Nobeyama Millimeter Array; \citet{imanishi07}) 
transitions of both species in Figure~2, after convolving
the ALMA images with the NMA beam ($7\farcs5 \times 5\farcs5$).
While the J=4--3 transitions of both species are concentrated near  
the eastern near-infrared nucleus with a slight extension to the west 
for HCO$^+$~(4--3), the distribution of the HCN~(1--0) and HCO$^+$~(1--0)
are different; the HCN~(1--0) emission is separated into 
two clumps in the east-west whereas the HCO$^+$~(1--0) emission is 
extended widely toward the western nucleus.

We label the six  
HCO$^+$(4--3) peaks in the eastern part of VV114 
as E0 -- E5 and the four detected in the western
part of VV114 as W0 -- W3 (Figure~1, Table~1).  
The HCO$^+$(4--3) and HCN~(4--3) emission peaks are spatially consistent  
for E0 -- E3.
The compact component E0 is unresolved with the current resolution, 
and the size upper limit is $< 200$~pc. 
HCN~(4--3) emission is not detected in the overlap region (W0 -- W2), 
where the high CO~(1--0) velocity dispersion and significant
methanol detection both suggest the presence of shocked gas 
(Saito et al. in preparation).

\section{Dense Gas and AGN /Starburst Activity}

\subsection{Relative Strengths of HCN~(4--3)  and HCO$^+$(4--3) and a Signature of AGN}

We present a comparison between the surface brightness of 
HCO$^+$(4--3) and HCN~(4--3) in Figure~3.  Although the statistics are limited, 
the three molecular clumps (E1, E2 and E3) show an increasing 
trend between $\Sigma_{\rm HCN}$ and $\Sigma_{\rm HCO^+}$. 
In contrast, the ratio between the beam averaged surface brightness of 
E0 is a factor of three higher than E1, E2 and E3;  
E0 is the only component that 
has HCN~(4--3) -- HCO$^+$(4--3) integrated flux ratio which is larger than unity
(HCN~(4--3)/HCO$^+$(4--3) = 1.6).  Gaussian fits to the
HCO$^+$(4--3) and HCN~(4--3) spectra at E0 give 
peak = 9.0~mJy, $\sigma = 123$~km~s$^{-1}$ (for HCN~(4--3)) and 
peak = 6.9~mJy, $\sigma = 93$~km~s$^{-1}$ (for HCO$^+$(4--3)).
Thus the HCN~(4--3) emission is not only brighter at E0, but it is also 
 broader than the gas traced in HCO$^+$(4--3), 
suggesting that the HCN~(4--3) and HCO$^+$(4--3) 
are tracing physically different gas at $< 200$~pc scales.
Such a high relative intensity of the HCN emission is 
possibly a signature of a buried AGN, as suggested by previous studies　
(e.g. \cite{kohno01}).

It has been known that the brightness of the 
HCN emission line is enhanced near the AGN compared
to star forming regions (\cite{kohno01}), 
with higher contrast in high J transitions 
(\cite{hsieh12}).  
Individual galaxies (e.g. NGC~1068, NGC~1097) have 
been studied extensively in high resolution 
(\cite{kohno03}, \cite{krips11}, \cite{hsieh12}), 
clearly revealing the over abundance of HCN emission 
near the Seyfert nucleus, through J = 1 to 4.
The exact reasoning for the enhanced intensity 
ratio is not clearly understood, and it could be due to gas excitation 
effects (e.g. density and temperature), intensity of the incident 
radiation field (e.g. PDR vs. XDR), IR pumping (e.g. \cite{garciaburillo06}), 
or other non-collisional 
excitation due to star formation or supernovae explosions
(see \cite{krips08} for a discussion).  
There is evidence 
suggesting the dominance of low density 
($< 10^{4.5}$~cm$^{-3}$) gas in a sample of AGNs \citep{krips08}, and hence
the difference in critical density is likely not the only 
reason for the difference in the relative abundance.  

\begin{figure}
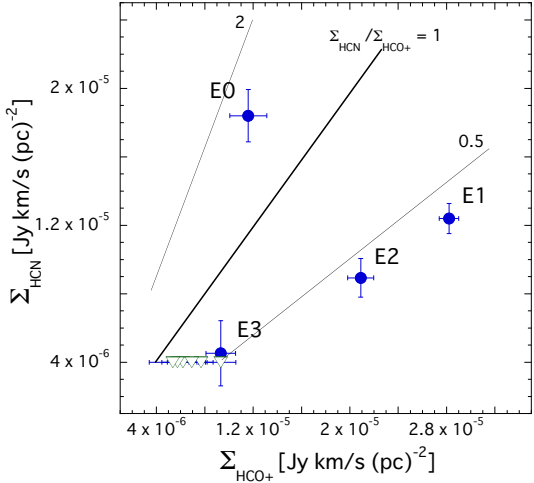

  \begin{center}
    \FigureFile(70mm,20mm){surface_densityV3.epsi}
  \end{center}
  \caption{Relation between the HCN~(4--3) and HCO$^+$(4--3) surface 
brightness for different
regions in VV114.  The triangles represent upper limits to the HCN~(4--3) surface brightness.  The solid lines are the surface brightness ratio of 0.5, 1 and 2.}
\end{figure}

Regardless of the exact physical origin of the higher relative 
intensity of the HCN~(4--3) emitting gas, 
the broad (FWHM = 290~km~s$^{-1}$) and compact ($< 200$~pc) 
unresolved source E0 is of significant interest, since 
it coincides with the region where past observations 
suggest the presence of a buried AGN.
We derive the upper limit to the dynamical mass by using
$M_{\rm dyn} = r \sigma^2/G$ (assuming an inclination of 
90 degrees for simplicity), 
where $r$ is the radius enclosing the emission region, 
$\sigma$ is the width of the HCN line, and 
$G$ is the gravitational constant.
The upper limit to the dynamical mass estimated from the 
line-width and the beam size is 
$\lesssim 4 \times 10^8$~M$_{\odot}$.
Since the HCN emission is generally believed to be optically thick,
we estimate the dense gas mass of the E0 component 
adopting the conversion factor provided in \citet{gao04}.  
Using the integrated intensity of HCN~(4--3) (see Section 3) and 
HCN~(1--0) = 7~Jy~km~s$^{-1}$ \citep{imanishi07}, we derive 
HCN(4--3)/(1--0) = 0.63.  This yields  
a dense gas mass of $\sim 8.1 \times 10^6$~M$_{\odot}$, 
hence $> 2$\% of the total mass is in dense molecular form and 
significant amount of dense gas is present in a very compact region.


Finally, we note that while these are evidences suggesting 
a compact AGN near the eastern 
nucleus of VV114, the 350~GHz -- 8.5~GHz flux ratio 
suggests the contrary.  The ratio is $1.2 \pm 0.1$ for E0, 
and $1.1 \pm 0.1$ for E1 and E2, using the 
the archival 8.5~GHz radio continuum image obtained 
from the VLA archive (beam size $\sim 0\farcs9$) 
and the 350~GHz image obtained from the ALMA observations.  
The 350~GHz continuum emission 
is also unresolved at E0, but E1 and E2 show resolved structure. 
If the dominant source of radio continuum 
emission is indeed due to hot plasma surrounding the AGN, then 
we expect this ratio to be higher near the putative AGN (i.e. E0), 
which is inconsistent with the current results 
and argues in favor of a
common physical origin (e.g. a massive starburst) in all three regions.
Higher resolution radio continuum imaging  
is necessary to understand the origin of the radio emission in E0.
 
\subsection{Extended Dense Gas Filament, Star Formation, and the Global Gas Conditions}

The average size of the clumps 
forming the filamentary structure (i.e. E1--E5, W0--W3) 
is $230 \pm 70$~pc, with an average dense gas mass of $\sim 10^{6}$~M$_{\odot}$
and a mean projected separation of $\sim 600$~pc.
We compare the distribution of the HCO$^+$(4--3) and star formation activity
traced in Pa$\alpha$ line in Figure~4.  
Spatial correspondence between HCO$^+$(4--3) and the brightest peaks of 
Pa$\alpha$ is generally seen.
Such a long filamentary dense gas structure and associated star 
formation are predicted along the colliding 
interface of two colliding galaxies \citep{saitoh09}, and the 
masses are also consistent with the massive star forming clumps 
predicted in simulations by \citet{teyssier10}.

Finally, we derive the global physical conditions of gas  
by comparing the total integrated HCO$^+$(4--3)/(1--0) 
and HCN(4--3)/(1--0) ratios to the 
results from radiative transfer modeling (RADEX; \cite{vandertak07}).   
The results are $n_{\rm H_2} = 10^5$ -- $10^6$~cm$^{-3}$ and T = 30 -- 500~K, 
assuming abundance ratios of 
[HCO$^+$]/[H$_2$] = $1.0 \times 10^{-9}$ \citep{irvine87} 
and [HCN]/[HCO$^+$] = 0.1 to 1 (to be consistent with M82; \cite{krips08}).  
Although the range in the derived temperature is too large to
be a meaningful constraint,
this suggests the presence of moderately dense gas averaged 
over the entire galaxy pair.  We caution here that these are 
average quantities which are derived without considering 
the difference in the spatial distribution between the J=4--3 and 1--0 transitions (see Figure~2).  
Higher angular resolution imaging 
of the J=1--0 transition is clearly needed in order to determine the spatial distribution of the
  physical properties.


\begin{figure}
  \begin{center}
    \FigureFile(70mm,15mm){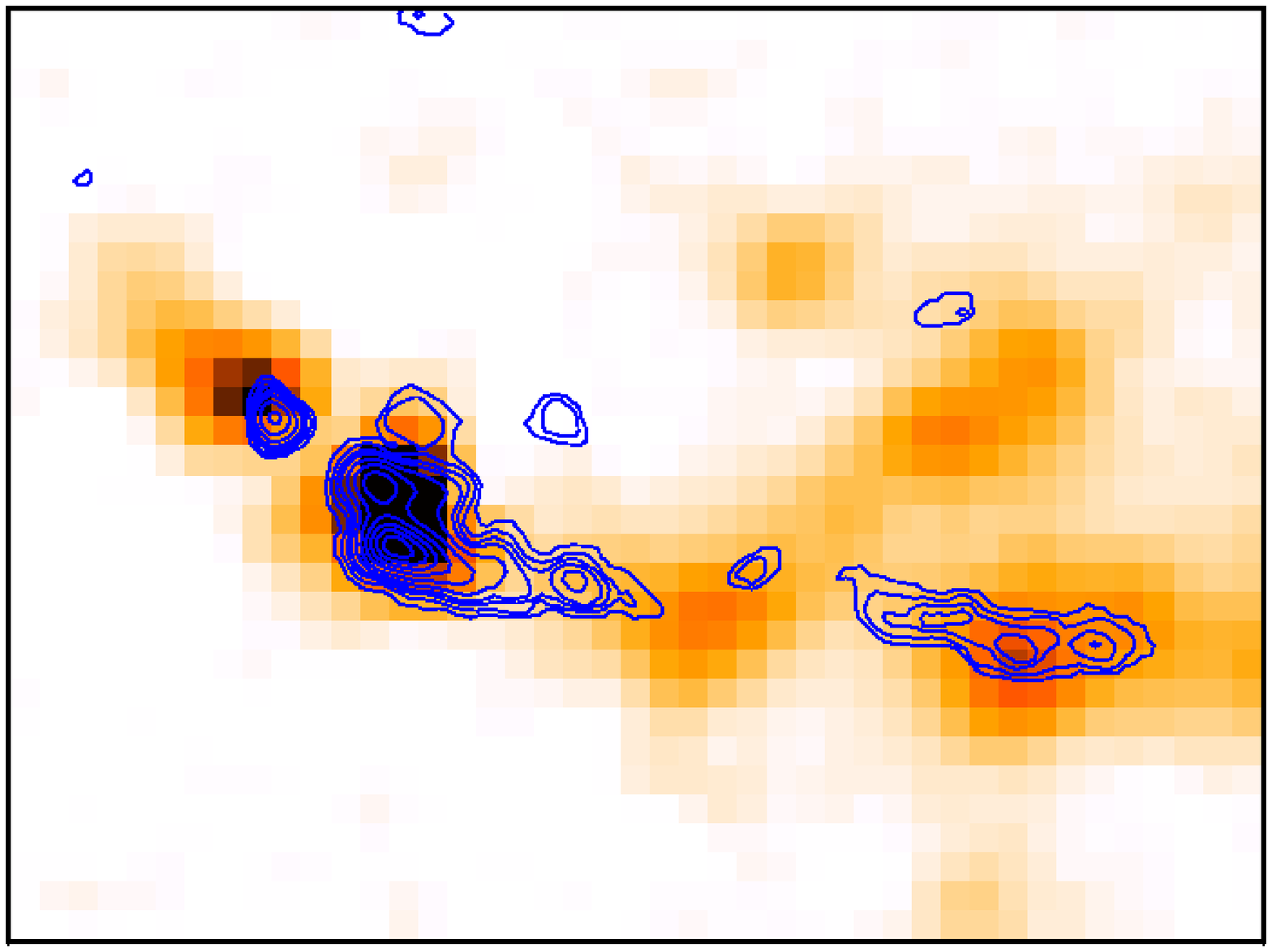}
  \end{center}
  \caption{HCO$^+$(4--3) emission overlaid on a Pa$\alpha$ image.  
The contour levels are the same as Figure~1. 
The Pa$\alpha$ map \citep{tateuchi12}
is obtained using the NIR camera ANIR \citep{motohara08} 
mounted on the University of 
Tokyo Atacama Observatory 1m telescope (miniTAO) \citep{minezaki10}.
A $\sim 1$~kpc long Pa$\alpha$ 
extension to the NE is also seen emanating 
from the putative AGN, 
which may be a signature of shock ionization, or 
star formation activity in the compressed gas along the AGN jet. 
}
\end{figure}

\begin{table}
\begin{center}
  \caption{Properties of the Molecular Clumps}\label{tab1}
  \begin{tabular}{lccccc}
  \hline           
  Source & $S_{\rm HCO^+}$~dv\footnotemark[1] & $S_{\rm HCN}$~dv\footnotemark[1]  & $T_{\rm HCO^+}$\footnotemark[2] & $T_{\rm HCN}$\footnotemark[2] \\ 
  & [Jy km/s] & [Jy km/s] & [K] & [K] \\ 
  \hline
  E0 & $1.12 \pm 0.15$ & $1.79 \pm 0.15$ & 0.40 & 0.57\\
  E1 & $7.40 \pm 0.21$ & $1.82 \pm 0.13$ & 1.33 & 0.63 \\
  E2 & $3.06 \pm 0.16$ & $0.64 \pm 0.08$ & 0.92 & 0.44\\
  E3 & $0.89 \pm 0.12$ & $0.11 \pm 0.05$ & 0.35 & 0.22\\
  E4 & $0.31 \pm 0.07$ &  -- & 0.30 & -- \\
  E5 & $0.20 \pm 0.04$ &  -- & 0.24 & -- \\
  W0 & $0.83 \pm 0.11$ &  -- & 0.37 & -- \\
  W1 & $0.88 \pm 0.11$ &  -- & 0.26 & -- \\
  W2 & $0.48 \pm 0.09$ &  -- & 0.32 & -- \\
  W3 & $0.12 \pm 0.04$ &  -- & 0.17 & -- \\
  \hline
\end{tabular}
\end{center}
\footnotemark[1] The integrated flux densities. \\
\footnotemark[2] The peak temperature of each molecular clump.  The error on each value is 0.05~K.

\end{table}

\section{Summary and Future Prospects}

We present $0\farcs4$ resolution HCN~(4--3) and HCO$^+$(4--3) observations 
toward a mid-stage IR bright merger VV114 obtained 
during cycle 0 program of ALMA.
For the first time, these new high-quality maps allow us to investigate 
the central regions of this merging LIRG at 200~pc resolution. 
We find that both the HCN~(4--3) and  HCO$^+$(4--3) emission in the 
eastern nucleus of VV~114 are compact ($< 200$~pc) and 
broad (290~km~s$^{-1}$ for HCN~(4--3)) with high HCN~(4--3)/HCO$^+$(4--3) 
ratio.  From the new ALMA observations along with 
past X-ray and NIR observations, we suggest the presence of an obscured AGN
in the eastern nucleus of VV114.
We also detect a 3--4~kpc long filament of dense gas, which is likely
 tracing the active star formation triggered by the ongoing merger. 
In a forthcoming paper, we will present a comprehensive modeling 
of VV~114 using our new $^{12}$CO~(1--0), $^{13}$CO~(1--0) and CO~(3--2) 
ALMA observations, as well as a chemical analysis of the nucleus 
and the overlap region of VV114 (Saito et al. in prep).

The authors thank the anonymous referee for comments that improved the contents of this paper.  This paper makes use of the following ALMA data: ADS/JAO.ALMA\#2011.0.00467.S. ALMA is a partnership of ESO (representing its member states), NSF (USA) and NINS (Japan), together with NRC (Canada) and NSC and ASIAA (Taiwan), in cooperation with the Republic of Chile. The Joint ALMA Observatory is operated by ESO, AUI/NRAO and NAOJ.



\bigskip


\end{document}